\documentclass[12pt]{iopart}
\usepackage{epsfig}
\usepackage{iopams}

\newcommand{\be}{\begin{equation}}
\newcommand{\ee}{\end{equation}}

\newcommand{\ba}{\begin{array}}
\newcommand{\ea}{\end{array}}
\newcommand{\bea}{\begin{eqnarray}}
\newcommand{\eea}{\end{eqnarray}}

\newcommand{\nn}{\nonumber \\}

\newtheorem{theorem}{Theorem}

\begin{document}

\title{A superintegrable finite oscillator in two dimensions with $SU(2)$ symmetry}

\author{Hiroshi Miki}
\ead{miki@amp.i.kyoto-u.ac.jp}
\address{Department of Applied Mathematics and Physics, Graduate School of Informatics, Kyoto University, Sakyo-Ku, Kyoto 606 8501, Japan}
\author{Sarah Post\footnote{The bulk of this research was performed while S.P. was a postdoctoral researcher at the Centre de Recherches Math\'{e}matiques, Universit\'{e} de Montr\'{e}al } }
\ead{spost@hawaii.edu}
\address{Department of Mathematics, University of Hawaii at Manoa\\
2565 McCarthy Mall, Honolulu HI 96822, USA}
\author{Luc Vinet}
\ead{luc.vinet@umontreal.ca}
\address{Centre de Recherches Math\'{e}matiques and D\'epartement de Physique , Universit\'{e} de Montr\'{e}al, P. O. Box 6128, Centre-ville Station, Montr\'{e}al (Qu\'{e}bec), H3C 3J7, Canada}
\author{Alexei Zhedanov}
\ead{zhedanov@fti.dn.ua}
\address{Donetsk Institute for Physics and Technology, Donetsk 83 114, Ukraine}

\begin{abstract} A superintegrable finite model of the quantum isotropic oscillator in two dimensions is introduced. It is defined on a uniform lattice of triangular shape. The constants of the motion for the model form an $SU(2)$ symmetry algebra. It is found that the dynamical difference eigenvalue equation can be written in terms of creation and annihilation operators. The wavefunctions of the Hamiltonian are expressed in terms of two known families of bivariate Krawtchouk polynomials; those of Rahman and those of Tratnik. These polynomials form bases for $SU(2)$ irreducible representations.  It is further shown that the pair of eigenvalue equations for each of these families are related to each other by an $SU(2)$ automorphism. A finite model of the anisotropic oscillator that has wavefunctions expressed in terms of the same Rahman polynomials
is also introduced.  In the continuum limit, when the number of grid points goes to infinity, standard two-dimensional harmonic oscillators are obtained. The analysis provides the $N\rightarrow \infty$ limit of the bivariate Krawtchouk polynomials as a product of one-variable Hermite polynomials.  
\end{abstract}

\section{Introduction}
There is a considerable body of knowledge on superintegrable models and remarkably, the recent years have witnessed significant advances in their classification. The harmonic oscillator is the prototype of superintegrable models. These are Hamiltonian systems which admit a number of conserved quantities greater than the dimension $d$ in which they are defined. They are called maximally superintegrable when this number is $2d-1$. In quantum mechanics, these constants of motion are typically required to be algebraically independent. To a large extent, the documented cases describe continuous systems and little is known in the discrete realm. It is the purpose of the present paper to present a finite analog of the quantum harmonic oscillator in two dimensions which is maximally superintegrable. 

There is by now a number of finite models of the quantum oscillator in one dimension. The most studied is based on the $SU(2)$ algebra \cite{Atakishiyev2005finite}. Generalizations that model the parabosonic oscillator have been recently developed with extended algebras \cite{Jafarov2011finite, Jafarov2011Hahn, Jafarov2012finite}. Let us recall the main features of the system associated with the $SU(2)$ algebra whose generators $J_0, J_\pm$ satisfy $\left[ J_0, J_{\pm}\right] =\pm J_\pm$ and $\left[J_+, J_-\right]=2J_0$ and whose standard $N+1$-dimensional irreducible representations may be taken to have basis vectors $| N,n\rangle$, $n=0, \ldots, N$, defined by $J_0|N, n\rangle=(n-N/2)|N,n \rangle$ with $N$ a non-negative integer related to the value of the Casimir operator. Upon making the identification 
\be H=J_0+\frac N 2 +\frac12, \qquad Q=\frac12(J_++J_-), \qquad P=\frac{i}{2}(J_+-J_-),\ee
the $SU(2)$ commutation relations are interpreted as the Heisenberg equations of motion of the oscillator: 
\be \left[ H, Q\right]=-iP, \qquad \left[H, P\right]=iQ.\ee
The wavefunctions of this finite oscillator are, naturally, the overlaps between the eigenstates of the position operator $Q$ and those of $H$, ie. $\{ |N, n\rangle\}$; they are hence given by rotation matrix elements known to be expressible in terms of the Krawtchouk polynomials \cite{koekoek2010askey} $K_n(x, \frac12, N)$ with $x=N/2 +q$ and $q\in \{ -N/2, ..., N/2\}$, the eigenvalues of $Q$. 
The Krawtchouk polynomials are given in terms of a hypergeometric series 
\be \label{K1d} K_n(x;p;N)=(-N)_n\sum_{j=0}^{N} \frac{(-n)_j(-x)_j}{j! (-N)_j}\left(\frac{1}{p}\right)^j \ee
and are orthogonal with respect to the binomial distribution
\be\label{weight1d} \left(\ba{c} N\\x\ea \right) p^x (1-p)^{N-x},\quad x=0,\cdots ,N.\ee
This entails a realization of the observables of this finite oscillator in terms of finite difference operators. It is known that $SU(2)$ contracts to the Heisenberg algebra. Hence, when $N\rightarrow \infty$  and with proper rescaling, the description of the standard (continuous) harmonic oscillator is recovered \cite{Atakishiyev2005finite} as the Krawtchouk polynomials tend to the Hermite polynomials $H_n(x)$ in the limit $N \rightarrow \infty$.

The Krawtchouk (or $SU(2)$) one-dimensional finite oscillator has been used to build two-dimensional models. A first approach was to take the direct product of two finite one-dimensional oscillators to obtain a two-dimensional system defined on a square grid of points and with $SU(2)\times SU(2)$ as dynamical algebra.  A second one \cite{Atakishiyev2001finiteI} was to use the isomorphism between $SO(4)$ and $SU(2)\oplus SU(2)$ to describe this finite two-dimensional oscillator in terms of discrete radial and angular coordinates. As remarked in Refs. \cite{Atakishiyev2005finite, Atakishiyev2001finiteI, Atakishiyev2001finiteII}, these models do not possess the $SU(2)$ symmetry algebra of the continuous two-dimensional quantum oscillator. This symmetry algebra (not to be confused with the dynamical algebras) is generated by the three constants of motion which make the continuous harmonic oscillator maximally superintegrable. The two-dimensional finite models constructed so far do not therefore share that property. We here  introduce a system that has such an $SU(2)$ symmetry. The model will be defined on a grid of triangular shape. 

Finite oscillator models have various uses e.g. in optical image processing and in signal analysis where only a finite number of eigenvalues exist \cite{ Atakishiyev1999continuous, Atakishiyev1997fractional}. Finite planar oscillators can be employed in particular to describe waveguides and pixellated screens. It is expected that preserving, at the discrete level, all the symmetries of the continuous oscillator model would prove advantageous in these applications. 

It is natural to think that two-variable generalizations of the Krawtchouk polynomials could provide an underpinning for interesting two-dimensional oscillator models. As a matter of fact, two families of such bivariate polynomials have been identified. Some 20 years ago, Tratnik introduced multivariable Racah polynomials depending on two parameters, that encompass multivariable Krawtchouk polynomials in their simplest case. Geronimo and Iliev \cite{ GI2010} subsequently showed that these polynomials are bispectral, i.e. that they obey, in the bivariate case, a pair of difference equations in addition to the recurrence relations associated with their orthogonality. Griffiths \cite{Griffiths71} in the early 70's and more recently Hoare and Rahman \cite{Hoare2008probabilistic} discussed a second family of bivariate Krawtchouk polynomials with four parameters in connection with a probabilistic model. (See also Ref. \cite{ Mizukawa2004hypergeometric}.) These were later extended to an arbitrary number of variables \cite{grunbaum2011system} and shown  to be also bispectral \cite{grunbaum2007rahman}. In contradistinction with the Tratnik version, the Krawtchouk polynomials of Rahman satisfy a nearest neighbor five-term recurrence relation \cite{grunbaum2010family}. They have been used to construct a spin lattice with remarkable quantum state transfer properties \cite{Miki2012quantum}. 

In view of the bispectrality of the two-variable Krawtchouk polynomials, it is of interest to inquire what is the quantum mechanics that their difference equations describe and to find the corresponding continuous dynamics in the limit $N\rightarrow \infty$. This last question involves obtaining the $N\rightarrow \infty$ limit of the bivariate Krawtchouk polynomials, an issue that seems not to have been resolved so far. 

The Rahman polynomials lead to the most direct interpretation. Remarkably, their $N\rightarrow \infty$ limit is found to be the product of two Hermite polynomials. This is obtained by observing that their difference equations factorize in terms of a pair of simple creation and annihilation operators which obey Heisenberg commutation relations. A finite model for the two-dimensional isotropic (or anisotropic) oscillator can then be constructed in the standard fashion. In terms of these creation and annihilation operators, the constants of motion take the same form as in the continuous situation and the symmetry algebra is therefore preserved in the discretization process. For the isotropic case, the Rahman polynomials are shown to form a basis for the irreducible representation spaces of $SU(2)$ They correspond in the continuum limit to wave functions that are separated in Cartesian coordinates. In view of the eigenvalue equations that they obey \cite{GI2010}, the Tratnik polynomials are seen to correspond to separation of variables in coordinates that are rotated relative to the ones associated with the Rahman polynomials. This is why the Tratnik polynomials depend essentially on one less parameter than those of Rahman; the additional parameter has been removed by the rotation.  It should be stressed  that the oscillator models based on these bivariate Krawtchouk polynomials are defined on a uniform grid of triangular shape. 

The remainder of the paper is organized as follows. In section 2, we present an $SU(2)$ invariant finite isotropic oscillator model, whose eigenfunctions are given in terms of the two-variable Krawtchouk polynomials of Hoare and Rahman, and we describe its continuum limit. Also, a finite anisotropic oscillator is introduced which is seen to be superintegrable for rational frequencies of the coupling constants.  In section 3, we show how the 
two-variable Krawtchouk polynomials of Tratnik can also be realized as eigenfunctions of the same isotropic Hamiltonian, though associated with a different integral of motion. Conclusions and an Appendix on the large $N$ limit of the trinomial distribution follow.

\section{Finite oscillator model and the two-variable Krawtchouk polynomials of  Hoare and Rahman}\label{secRahman}
We first begin with the two-variable Krawtchouk polynomials introduced by Hoare and Rahman \cite{Hoare2008probabilistic} which are a special case of the Aomoto-Gelfand hypergeometric function \cite{Mizukawa2004hypergeometric}:   
\be \label{Kmn} K_{m,n}^N=\sum_{0\leq i+j+k+\ell\leq N}\frac{(-m)_{i+j}(-n)_{k+\ell}(-x)_{i+k}(-y)_{j+\ell}}{i!j!k!\ell! (-N)_{i+j+k+\ell}}u_1^iv_1^ju_2^kv_2^\ell, \ee
with parameters
\be\label{us} \ba{ll} u_1=\frac{(p_1+p_2)(p_1+p_3)}{p_1\sum p_i}, &u_2=\frac{(p_1+p_2)(p_2+p_4)}{p_2\sum p_i}\\
v_1=\frac{(p_1+p_3)(p_3+p_4)}{p_3\sum p_i}, & v_2=\frac{(p_2+p_4)(p_3+p_4)}{p_4\sum p_i}.\ea\ee
It should be noted that the polynomials depend essentially on three independent parameters. This fact can be observed in terms of the $p_i$ by writing
\be u_1=\frac{(1+p_2/p_1)(1+p_3/p_1)}{1+p_2/p_1+p_3/p_1+p_4/p_1}\ee
and equivalently for $u_2, v_1$ and $v_2$.

With $m$ and $n$ nonnegative integers such that $0\le m+n \le N$, the polynomials $K_{m,n}^N$ are defined on a uniform lattice of a triangular shape, i.e. $\{ (x,y) \in \mathbb{Z}_{\ge 0}^2~|~0\leq x+y\leq N\}$ and are orthogonal with respect to the two-variable generalization of the weight \eref{weight1d} given by the trinomial distribution 
\be\label{Kraw weight} \omega(x,y)=\left(\! \ba{c} N\\ x,y\ea\!\right)\eta_1^x\eta_2^y(1-\eta_1-\eta_2)^{N-x-y},\ee
with 
\bea \eta_1&&= \frac{p_1p_2(p_1+p_2+p_3+p_4)}{(p_1+p_2)(p_1+p_3)(p_2+p_4)},  \nonumber \\
 \eta_2&&=\frac{p_3p_4(p_1+p_2+p_3+p_4)}{(p_2+p_4)(p_3+p_4)(p_1+p_3)}.\eea
It should be noted that the constants $u_i,v_i,\eta_i$ satisfy the following three functional relations \cite{grunbaum2010family}:
\begin{eqnarray}
&&u_1\eta_1+v_1 \eta_2=1, \nonumber \\
&&u_2\eta_1+v_2 \eta_2=1, \nonumber \\
&&u_1u_2\eta_1 +v_1v_2 \eta_2=1,
\end{eqnarray}
which also imply that the polynomials $K_{m,n}^N$ depend essentially on three parameters.

One of the specific features which the polynomials $K_{m,n}^N$ possess is that they obey a nearest neighbor difference equation in $x,y$  \cite{grunbaum2007rahman} (and also some contiguity relations in $m,n$ because of their duality):
\begin{eqnarray}
&&[(p_1+p_3)m-(p_2+p_4)n]K_{m,n}^N  \nonumber \\
&&=(N-x-y)\Biggl\{ \frac{p_1p_2(p_3+p_4)\sum p_i}{p_1+p_2}\Delta _x-\frac{p_3p_4(p_1+p_2)\sum p_i}{p_3+p_4}\Delta_y  \Biggr\} K_{m,n}^N \nonumber \\
&&\quad +x\frac{p_1p_4-p_2p_3}{p_1+p_2}\Delta_{-x} K_{m,n}^N -y\frac{p_1p_4-p_2p_3}{p_3+p_4}\Delta_{-y}K_{m,n}^N , \label{5term}
\end{eqnarray}
where $\Delta$ is a difference operator defined by $\Delta_{\pm l_1,\pm l_2,\cdots }f(l_1,l_2,\cdots )=f(l_1\pm 1,l_2\pm 1,\cdots )-f(l_1,l_2,\cdots )$. Recently, it was shown \cite{Iliev2012rahman,grunbaum2007rahman} that 
the polynomials $K_{m,n}^N$ additionally satisfy a seven point difference equation, which is equivalent to the fact that the polynomials $K_{m,n}^N$ are eigenfunctions of the following difference operators: 
\begin{eqnarray}
\Lambda _1^N =&& \frac{(N-x-y)p_1p_3\sum p_i}{(p_1+p_3)(p_1p_4-p_2p_3)} \left( \frac{p_2 }{p_1+p_2} \Delta _x -\frac{p_4}{p_3+p_4}\Delta_y \right) \nonumber \\ 
                  &&+\frac{p_1p_4-p_2p_3}{(p_1+p_2)(p_3+p_4)}\left( x\frac{p_3}{p_1+p_3} \Delta_{-x} -y\frac{p_1}{p_1+p_3} \Delta_{-y} \right) \nonumber \\
                  &&-x\frac{p_3p_4}{(p_1+p_3)(p_3+p_4)}\Delta_{-x,y}-y\frac{p_1p_2}{(p_1+p_2)(p_1+p_3)}\Delta_{x,-y} \nonumber \\ 
\Lambda _2^N =&& \frac{(N-x-y)p_2p_4\sum p_i}{(p_2+p_4)(p_1p_4-p_2p_3)} \left( -\frac{p_1 }{p_1+p_2} \Delta _x +\frac{p_3}{p_3+p_4}\Delta_y \right) \nonumber \\ 
                  &&\frac{p_1p_4-p_2p_3}{(p_1+p_2)(p_3+p_4)}\left( -x\frac{p_4}{p_2+p_4} \Delta_{-x} +y\frac{p_2}{p_2+p_4} \Delta_{-y} \right) \nonumber \\
                  &&-x\frac{p_3p_4}{(p_2+p_4)(p_3+p_4)}\Delta_{-x,y}-y\frac{p_1p_2}{(p_1+p_2)(p_2+p_4)}\Delta_{x,-y}. \label{l1l2}
\end{eqnarray}
The eigenvalue equations read:
\bea \Lambda_1^N K_{m,n}^N=mK_{m,n}^N, \nonumber \\
        \Lambda_2^N K_{m,n}^N=nK_{m,n}^N.\eea
As it will later become important, the sum  $\Lambda_1^N +\Lambda_2^N$ depends only on the two parameters $\eta_1$ and $\eta_2$.        
 
 A different  combination of the operators \eref{l1l2} eliminates the non-nearest neighbor terms involving $\Delta_{x,-y},~\Delta_{-x,y}$  and the difference equation \eref{5term} recovered.
Moreover, one can easily find that $p_2(p_1+p_3)\Lambda_1^N+ p_1(p_2+p_4)\Lambda_2^N$ and  $p_4(p_1+p_3)\Lambda_1^N+ p_3(p_2+p_4)\Lambda_2^N$ give the ``recurrence'' relations for the polynomials $K_{m,m}^N$:
\begin{eqnarray}
\fl (p_2(p_1+p_3)m+p_1(p_2+p_4)n)K_{m,n}^N(x,y) \nonumber \\
=-\Biggl[ (N-x-y)\frac{p_1p_2\sum p_i }{p_1+p_2} \Delta_x +x\frac{(p_1p_4-p_2p_3)^2}{(p_1+p_2)(p_3+p_4)}\Delta_{-x} \nonumber  \\
\quad~ +x\frac{p_3p_4(p_1+p_2)}{p_3+p_4}\Delta_{-x,y}+yp_1p_2\Delta_{x,-y}\Biggr] K_{m,n}^N, \nonumber \\
\fl (p_4(p_1+p_3)m+p_3(p_2+p_4)n)K_{m,n}^N(x,y) \nonumber \\
=-\Biggl[ (N-x-y)\frac{p_3p_4\sum p_i }{p_3+p_4} \Delta_y +y\frac{(p_1p_4-p_2p_3)^2}{(p_1+p_2)(p_3+p_4)}\Delta_{-y} \nonumber  \\
\quad +xp_3p_4\Delta_{-x,y}+y\frac{p_1p_2(p_3+p_4)}{p_1+p_2}\Delta_{x,-y}\Biggr] K_{m,n}^N. \label{rec}
\end{eqnarray}
If we set $K_{m,n}=0$ when $x$ or $y$ is a negative integer and given that $K_{m,n}=1$ at $x=y=0$ from \eref{Kmn},  we can determine all the $K_{m,n}^N(x,y)$ from \eref{rec}. In that sense, the operators $\Lambda_1^N$ and $\Lambda_2^N$ fully characterize the Rahman polynomials $K_{m,n}^N$ defined on the grid points such that $0\le x+y \le N$ as their joint eigenfunctions.

\subsection{The $SU(2)$ invariant isotropic Hamiltonian }      

In this section, we first consider the operators defined by
\begin{eqnarray}
  A^{(R)}_-&&=\frac{p_1p_2p_3p_4\sum p_i}{(p_1+p_3)(p_1p_4-p_2p_3)}\left(\frac{1}{p_4(p_1+p_2)}\Delta_x-\frac{1}{p_2(p_3+p_4)}\Delta_y \right), \nonumber \\
 A^{(L)}_-&&=\frac{p_1p_2p_3p_4\sum p_i}{(p_2+p_4)(p_1p_4-p_2p_3)}\left(\frac{-1}{p_3(p_1+p_2)}\Delta_x+ \frac{1}{p_1(p_3+p_4)}\Delta_y \right), \nonumber \\
A_+^{(R,N)}&&=\frac{p_1p_4-p_2p_3}{p_1+p_2+p_3+p_4}\left(\frac x{p_1}T_x^{-1}-\frac{y}{p_3}T_y^{-1}\right)+(N+1-x-y) \nonumber \\
A_+^{(L,N)}&&=\frac{p_1p_4-p_2p_3}{p_1+p_2+p_3+p_4}\left(\frac{-x}{p_2}T_x^{-1}+\frac{y}{p_4}T_y^{-1}\right)+(N+1-x-y), \label{As}
\end{eqnarray}
where $T$ is the shift operator defined by $T_lf(l)=f(l+1)$. One can easily verify that these operators satisfy a shifted form of the Heisenberg algebra relations:
\bea A_-^{(i)}A_+^{(i,N)}-A_+^{(i,N-1)}A_-^{(i)}=1, \qquad i=R,L \nonumber \\
        A_-^{(R)}A_-^{(L)}-A_-^{(L)}A_-^{(R)}=0, \nonumber \\
        A_+^{(R,N+1)}A_+^{(L,N)}-A_+^{(L,N+1)}A_+^{(R,N)}=0, \nonumber \\
        A_-^{(R)}A_+^{(L,N)}-A_+^{(L,N-1)}A_-^{(R)}=0, \nonumber \\
        A_+^{(R,N+1)}A_-^{(L)}-A_-^{(L)}A_+^{(R,N)}=0, \label{ha}
        \eea
and provide the factorizations of the operators $\Lambda _1^N$ and $\Lambda _2^N$:
\be \Lambda_1^N=A_+^{(R, N-1)}A_-^{(R)}, \qquad  \Lambda_2^N=A_+^{(L, N-1)}A_-^{(L)} . \label{fa}\ee
Let us consider the polynomials $A^{(R)}_-K_{m,n}^N$. It is straightforward to  check from \eref{ha} that 
\begin{eqnarray}
&&\Lambda _1^{N-1}(A^{(R)}_-K_{m,n}^N) = (m-1)A^{(R)}_-K_{m,n}^N, \nonumber \\
&&\Lambda _2^{N-1}(A^{(R)}_-K_{m,n}^N) = n A^{(R)}_-K_{m,n}^N.
\end{eqnarray}
This means that $A^{(R)}_-K_{m,n}^N$ are also the Rahman polynomials $K_{m-1,n}^{N-1}$ up to a multiplicative constant, which is determined by evaluating $\Lambda _-^{(R)}K_{m,n}^N$ at $(x,y)=(0,0)$. We can thus obtain the following four relations:
\bea
A^{(R)}_- K_{m,n}^N=\frac{m}{N}K_{m-1,n}^{N-1},\qquad 
A^{(L)}_- K_{m,n}^N=\frac{n}{N}K_{m,n-1}^{N-1}, \nonumber \\
A^{(R,N)}_+K_{m,n}^{N}=(N+1) K_{m+1,n}^{N+1},\qquad 
A^{(L,N)}_+K_{m,n}^{N}=(N+1) K_{m,n+1}^{N+1}.\eea
Hence, the operators \eref{As} provide ladder operators for the Rahman polynomials \eref{Kmn}.

%

Now let us go back to the relations \eref{ha}. From these, it is clear that the Hamiltonian 
\be\label{H} \mathfrak{h}_{iso}=\Lambda_1^N+\Lambda_2^N \ee
admits the following integrals of the motion 
\bea  J_X=\frac{1}{2}\left( A_+^{(R,N-1)}A_-^{(L)}+A_+^{(L,N-1)}A_-^{(R)} \right), \nonumber \\
J_Y=\frac{i}{2}\left(A_+^{(R,N-1)}A_-^{(L)}-A_+^{(L,N-1)}A_-^{(R)}\right), \nonumber \\
 J_Z=\frac{1}{2}\left(A_+^{(R,N-1)}A_-^{(R)}-A_+^{(L,N-1)}A_-^{(L)}\right). \label{Z}\eea
The conserved quantities \eref{Z} form a basis for an $SU(2)$ algebra as they satisfy 
\be [J_X,J_Y]=iJ_Z, \quad [J_Y, J_Z]=iJ_X, \quad [J_Z, J_X]=iJ_Y\ee
and hence the Hamiltonian $\mathfrak{h}_{iso}$ of \eref{H} is $SU(2)$ invariant. It is readily seen that the Casimir operator
\be
Q=J_X^2+J_Y^2+J_Z^2
\ee 
takes the form 
\be
Q=j(j+1) ,\quad  j=\frac{m+n}{2}.
\ee
 It hence follows that the  Rahman polynomials \eref{Kmn} with fixed $j=(m+n)/2$ form a basis for the $2j+1$-dimensional irreducible representations of $SU(2)$. They diagonalize the operators $\Lambda_1^N, \Lambda_2^N$ or equivalently the commuting pair $\mathfrak{h}_{iso}$ and $J_Z$ with eigenvalues
 \be \mathfrak{h}_{iso}K_{m,n}^N=(m+n)K_{m,n}^N, \qquad J_ZK_{m,n}^N=\frac12(m-n)K_{m,n}^N.\ee

\subsection{An anisotropic Hamiltonian}
Anisotropic oscillators can also be constructed in the standard way using the ladder operators \eref{As} and the ``number'' operators $\Lambda_1^N$ and $\Lambda _2^N$ factorized in \eref{fa}.
Systems with
\be\label{Hani} \mathfrak{h}_{aniso}=\omega_1^2 \Lambda_1^N+\omega_2^2\Lambda_2^N\ee
as Hamiltonians will be integrable for arbitrary $\omega_{1}$ and $\omega_2$ and superintegrable when the ratio of the frequencies is rational. In each case, the eigenfunctions for the pair $\mathfrak{h}_{aniso}$ and either $\Lambda_1^N$ or $\Lambda_2^N$ are the Rahman polynomials. It is important to note that whereas $\mathfrak{h}_{iso}$ depends essentially only on two parameters $\eta_1$ and $\eta_2$, the anisotropic Hamiltonian depends on all three functionally independent parameters. 

An interesting observation in this context is that the spin lattice Hamiltonian introduced in \cite{Miki2012quantum} is generically associated with an anisotropic oscillator. In fact, a Hamiltonian of the form \eref{Hani} will have nearest-neighbor interactions only if the relation
\begin{equation}
\left( \frac{\omega _1}{\omega _2}\right)^2 = -\frac{p_1+p_3}{p_2+p_4} 
\end{equation} 
is verified.

\subsection{Continuum limit}
In this subsection, we consider the $N\rightarrow \infty$ limit of the two-variable Krawtchouk polynomials \eref{Kmn}. The limit will be obtained by  the following change of variables 
\be x=N\eta_1+\sqrt{N}\left(c_1s +c_2t\right), \qquad y=N\eta_2+\sqrt{N}\left(c_3s +c_4t\right),\label{xy}\ee
with 
\bea c_1=\frac{-p_2}{(p_1+p_2)(p_1+p_3)}\sqrt{\frac{2(p_1+p_2+p_3+p_4)p_1p_3}{p_2+p_4}}  \nonumber \\ c_2=\frac{p_1}{(p_1+p_2)(p_2+p_4)}\sqrt{\frac{2(p_1+p_2+p_3+p_4)p_2p_4}{p_1+p_3}} \nonumber \\
c_3=\frac{p_4}{(p_3+p_4)(p_1+p_3)}\sqrt{\frac{2(p_1+p_2+p_3+p_4)p_2p_4}{p_1+p_3}}\nonumber  \\ c_4=\frac{-p_3}{(p_3+p_4)(p_2+p_4)}\sqrt{\frac{2(p_1+p_2+p_3+p_4)p_2p_4}{p_1+p_3}}.\label{cs}\eea
Note that as $N\rightarrow \infty $, the range of $s$ and $t$ becomes the whole real line.

The operators $\Lambda_1^N, \, \Lambda_2^N$ have the following limits
\be\label{L12} \lim_{N\rightarrow \infty} \Lambda_1^N=   -\frac12\partial_s^2+s\partial_s,\qquad  \lim_{N\rightarrow \infty} \Lambda_2^N=-\frac12 \partial_t^2+t\partial_t .\ee
Thus, in the limit $N\rightarrow \infty$, the Hamiltonian $\mathfrak{h}_{iso}$ given in \eref{H} tends to the Hamiltonian of a two-dimensional oscillator 
\be \lim_{N\rightarrow \infty} \mathfrak{h}_{iso}=-\frac12\left(\partial_s^2+\partial_t^2\right)+s\partial_s+t\partial_t\ee
 which has been conjugated by the ground state and with the coupling constants absorbed into the variables.

While the operators $\Lambda_1^N$ and $\Lambda_2^N$ do not change the quantum numbers $m,n$ and $N$, the ladder operators do and so the the normalization of the polynomials (see below) affects the limit of these operators. In order to find the proper choice of normalization, note that in the large $N$ limit the trinomial distribution tends to the two variable Gaussian (for a proof see \ref{limitweight}) with the following coefficients
\be\fl \omega (x,y)= \left(\! \ba{c} N\\ x,y\ea\!\right)\eta_1^x\eta_2^y(1-\eta_1-\eta_2)^{N-x-y}= \frac{e^{-s^2-t^2} }{2\pi N\sqrt{\eta_1\eta_2(1-\eta_2-\eta_2)}}+\mathcal{O}\left(N^{-\frac32}\right).\ee 
Note that the constants in the change of variables \eref{cs} satisfy 
\[c_1c_4-c_2c_3=2\sqrt{\eta_1\eta_2(1-\eta_1-\eta_2)}\]
 and so the unit of area, as $x$ and $y$ are shifted by 1, is  $(2N\sqrt{\eta_1\eta_2(1-\eta_1-\eta_2)})^{-1}.$
Thus, 
\be \fl \lim_{N\rightarrow \infty} \sum_{0\le x+y\le N} F(x,y)\omega (x,y)=\frac{1}{\pi}\int_{-\infty}^{\infty}\int_{-\infty}^{\infty} \left[\lim _{N\rightarrow \infty} F\left(x(s,y), y(s,t)\right)\right] e^{-s^2-t^2}dsdt.\ee

The normalization of the Rahman polynomials given in \cite{grunbaum2010family} is
\be 
\sum_{0\le x+y\le N} \omega(x,y) K_{m_1,n_1}^N K_{m_2,n_2}^N(x,y)=I_{m_1,m_2}^{n_1,n_2}
\ee
with
\be
I_{m_1,m_2}^{n_1,n_2}=\delta_{m_1,m_2}\delta_{n_1,n_2} \frac{m_1!n_1!(N-m_1-n_1)!(p_1p_4-p_2p_3)^{2(m_1+n_1)}}{N! (p_1p_3(p_2+p_4))^{m_1}(p_2p_4(p_1+p_3))^{n_1}(\sum p_k)^{m_1+n_1}}  \label{normI}.
\ee
This suggests the following redefinition:
\begin{eqnarray}
&&\widehat{K}_{m,n}^N(x,y)=\alpha _{m,n}^NK_{m,n}^N(x,y) \nonumber \\
&&\alpha _{m,n}^N=\frac{(-\sqrt{2\sum p_i})^{m+n}}{(p_1p_4-p_2p_3)^{m+n}} \sqrt{\frac{N! \left(p_1p_3(p_2+p_4)\right)^m\left(p_2p_4(p_1+p_3)\right)^n}{(N-m-n)!}}. \label{widehatK}
\end{eqnarray}
With this normalization, the orthogonality relation becomes
\bea   
 \sum_{0\le x+y\le N} \omega (x,y)\widehat{K}^N_{m, n}\widehat{K}^N_{m', n'}=2^{m+n}n!m!\delta_{n, n'}\delta_{m, m'},\eea
which implies, in the limit
\bea   \frac{1}{\pi}\int_{-\infty}^{\infty}\int_{-\infty}^{\infty}e^{-s^2-t^2}\left(\lim_{N\rightarrow \infty} \widehat{K}^N_{m, n}\right)^2dsdt=2^{m+n}n!m!\label{norm}.\eea
Thus, the functions $\lim_{N\rightarrow \infty} \widehat{K}^N_{m, n}$ satisfy the eigenvalue equation
\be  \left[ -\frac12\partial_s^2+s\partial_s\right]  \left( \lim_{N\rightarrow \infty} \widehat{K}^N_{m, n} \right)=m \left( \lim_{N\rightarrow \infty} \widehat{K}^N_{m, n} \right) \ee
and a similar one in $t$. They are moreover square integrable in the plane with respect to the measure $ e^{-s^2-t^2}$ and are hence polynomial. The normalization constant has been chosen to agree with a product of Hermite polynomials and so the limit satisfies
\be \lim_{N\rightarrow \infty} \widehat{K}^N_{m, n} =\epsilon H_m(s)H_n(t), \qquad \epsilon^2=1.\label{Klimitfinal}\ee

Using the proper normalization, the action of the ladder operators becomes
\bea -\frac{\sqrt{2p_1p_3(p_1+p_2+p_3+p_4)(p_2+p_4)}}{(p_1p_4-p_2p_3)\sqrt{N+1}} A_+^{(m,N)}\widehat{K}^{N}_{m,n}=\widehat{K}^{N+1}_{m+1,n},  \nonumber \\ \frac{-\sqrt{N}(p_1p_4-p_2p_3)}{\sqrt{2p_1p_3(p_1+p_2+p_3+p_4)(p_2+p_4)}} A_-^{(m,N)}\widehat{K}^N_{m,n}=m\widehat{K}^{N-1}_{m-1,n},\eea
and similarly for the operators $A_\pm^n.$ 
As expected, the limit of these scaled operators becomes the ladder operators for the Hermite polynomials
\bea a_+^m\equiv \lim_{N\rightarrow \infty} -\frac{\sqrt{2p_1p_3(p_1+p_2+p_3+p_4)(p_2+p_4)}}{\sqrt{N+1} (p_1p_4-p_2p_3)}A_+^{(m,N)}
=2s-\partial_s, \nonumber \\
        a_-^m\equiv \lim_{N\rightarrow \infty} \frac{-\sqrt{N}(p_1p_4-p_2p_3)}{\sqrt{2p_1p_3(p_1+p_2+p_3+p_4)(p_2+p_4)}}A_-^{(m,N)}=\frac12\partial_s, \nonumber \\
        a_+^n\equiv \lim_{N\rightarrow \infty} -\frac{\sqrt{2p_2p_4(p_1+p_2+p_3+p_4)(p_1+p_3)}}{\sqrt{N+1}(p_1p_4-p_2p_3)}A_+^{(n,N)}=2t-\partial_t, \nonumber \\
        a_-^n\equiv \lim_{N\rightarrow \infty} \frac{-\sqrt{N}(p_1p_4-p_2p_3)}{\sqrt{2p_2p_4(p_1+p_2+p_3+p_4)(p_1+p_3)}}A_-^{(n,N)} =\frac12\partial_t .\eea
Since these ladder operators correspond exactly to those of Hermite polynomials,  the sign in \eref{Klimitfinal} does not depend on either $m$ or $n$. Furthermore, since $\widehat{K}_{0,0}^0=H_0(s)H_0(t)=1, $ the sign in \eref{Klimitfinal} is exactly $\epsilon=1.$

\section{Two-variable Krawtchouk polynomials of Tratnik}
A different set of commuting operators in the enveloping algebra of \eref{As} lead to the version of two-variable Krawtchouk polynomials defined by Tratnik \cite{Trat1989, Trat1991}. Beginning with the observation that the isotropic Hamiltonian $\mathfrak{h}_{iso}$ depends only on the two parameters $\eta_1$ and $\eta_2$, it is also possible to construct another operator written in terms of the creation and annihilation operators \eref{As}, given \eref{fa}, that  depends only on these two parameters. The set of commuting operators is given by 
\bea  \mathcal{L}_1=\mathfrak{h}_{iso}=\Lambda_1^N+\Lambda_2^N \nonumber \\
		\mathcal{L}_2=\ell_2\left[\frac{p_2p_4 A_+^{(L,N)}A_-^{(R)}}{(p_2+p_4)}+\frac{p_1p_3A_+^{(R,N)}A_-^{(L)}}{(p_1+p_3)}+\frac{p_1p_4\Lambda_1^N}{(p_2+p_4)}+\frac{p_2p_3\Lambda_2^N}{(p_1+p_3)}\right],\eea
		where
		\[
		\ell_2=\left( \frac{p_1p_4}{p_2+p_4}+\frac{p_2p_3}{p_1+p_3}\right)^{-1}.\]
Thus, the joint eigenfunctions of these operators will depend essentially only on two parameters. These eigenfunctions are the two-variable Krawtchouk polynomials of Tratnik as the operators $\mathcal{L}_1$ and $\mathcal{L}_2$  are exactly those given in Geronimo and Illiev \cite{GI2010}.  Since  $\mathcal{L}_2$ is a difference operator in $y$ alone, the two-variable Krawtchouk polynomials can be written as imbricated ordinary Krawtchouk polynomials
\be  \label{TratK} K_{2}(\vec{n}; x,y; \vec{\mathfrak{p}}; N)=\frac{k_{n_1}\left(x, \mathfrak{p}_1, N-n_2\right)}{(-N)_{n_1+n_2}} \cdot k_{n_2}\left(y; \frac{\mathfrak{p}_2}{1-\mathfrak{p}_1}; N-x\right),   \ee
where 
\begin{eqnarray}
\mathfrak{p}_1&=&\eta_1=\frac{p_1p_2\sum p_i}{(p_1+p_2)(p_1+p_3)(p_2+p_4)}, \nonumber \\ 
\mathfrak{p}_2&=&\eta_2=\frac{p_3p_4\sum p_i}{(p_3+p_4)(p_1+p_3)(p_2+p_4)}.\label{frakp}
 \end{eqnarray}
 The eigenvalue equations are
 \bea \mathcal{L}_1K_{2}(\vec{n}; x,y; \vec{\mathfrak{p}}; N)=(n_1+n_2)K_{2}(\vec{n}; x,y; \vec{\mathfrak{p}}; N), \\
 \mathcal{L}_2K_{2}(\vec{n}; x,y; \vec{\mathfrak{p}}; N)=n_2K_{2}(\vec{n}; x,y; \vec{\mathfrak{p}}; N).\eea
While both the Tratnik and Rahman polynomials diagonalize the isotropic Hamiltonian, the Rahman polynomials depend on an additional parameter. The reason for this is that the two families of polynomials each diagonalize a pair of operators which are related by an $SU(2)$ automorphism.

Recall from Section \ref{secRahman} that the Rahman polynomials are eigenfunctions for the isotropic oscillator Hamiltonian  $\mathfrak{h}_{iso}$ and the operator $J_Z$. The Tratnik polynomials are also eigenfunctions for $\mathfrak{h}_{iso}$ as well as the operator 
\be \label{frakk} \mathfrak{k}=\frac12\mathcal{L}_1-\mathcal{L}_2,\ee
with eigenvalues 
\be \mathfrak{k}K_{2}(\vec{n}; x,y; \vec{\mathfrak{p}}; N) =\frac12(n_1-n_2) K_{2}(\vec{n}; x,y; \vec{\mathfrak{p}}; N).\ee
Because of their eigenvalues, it is natural to expect that the operators $J_Z$ and $\mathfrak{k}$ are related and indeed $\mathfrak{k}$ can be expressed as
\be \mathfrak{k}=aJ_x+bJ_y+cJ_z,\qquad a^2+b^2+c^2=1\ee
with 
\bea a=\ell_2\left(\frac{p_2p_4}{p_2+p_4}+\frac{p_1p_3}{p_1+p_3}\right)\\
b=i\ell_2\left(\frac{p_2p_4}{p_2+p_4}-\frac{p_1p_3}{p_1+p_3}\right)\\
c=2\ell_2\frac{p_2p_3}{p_1+p_3}-1.\eea
This of course corresponds to the observation that $\mathfrak{k}$ is obtained from $J_Z$ by a rotation.

The Tratnik polynomials \eref{TratK} also tend to products of Hermite polynomials in the continuum limit as their imbricated expression suggests. The limit of the eigenvalue operators for these polynomials takes the form 
\begin{eqnarray}
 \lim_{N\rightarrow \infty} \mathcal{L}_1&&=-\frac12\left(\partial_s^2+\partial_t^2\right)+s\partial_s+t\partial_t, \nonumber   \\
 \lim_{N\rightarrow \infty}\mathcal{L}_2 
&&=\ell_2\left[\frac{p_1p_4}{2(p_1+p_2)(p_2+p_4)}\frac{\partial^2}{\partial s^2}+ \frac{p_2p_3}{2(p_1+p_2)(p_1+p_3)}\frac{\partial^2}{\partial t^2}\nonumber\right. \\
&&-\left(\frac{p_1p_4s}{(p_1+p_2)(p_2+p_4)}+t\sqrt{\frac{p_1p_2p_3p_4}{(p_1+p_2)^2(p_1+p_3)(p_2+p_4)}}\right)\frac{\partial }{\partial s}\nonumber \\ 
&&-\left(\frac{p_2p_3t}{(p_1+p_2)(p_1+p_3)}+s\sqrt{\frac{p_1p_2p_3p_4}{(p_1+p_2)^2(p_1+p_3)(p_2+p_4)}}\right)\frac{\partial }{\partial t} \nonumber \\
&&\left.+\sqrt{\frac{p_1p_2p_3p_4}{(p_1+p_2)^2(p_1+p_3)(p_2+p_4)}}\frac{\partial^2}{\partial t \partial s}\right] .
\end{eqnarray}
As expected, the operator $\mathcal{L}_2$ can be rotated into a linear combination of two eigenvalue operators for Hermite polynomials in the new coordinates.

\section{Concluding Remarks}
We have introduced and described a finite oscillator model in two dimensions which has the same symmetry as its continuum limit, the standard two-dimensional isotropic harmonic oscillator. This finite model is constructed using eigenvalue operators for two-variable Krawtchouk polynomials. These operators were shown to factorize into ladder operators that obey Heisenberg-like commutation relations. The $SU(2)$ symmetry algebra of the Hamiltonian is then obtained \`{a} la Schwinger using these raising and lowering operators and thus making the superintegrablity of the model manifest. As a consequence, the bivariate Krawtchouk polynomials were found to form bases for irreducible representations of $SU(2)$.

The Hamiltonian can be diagonalized by both the two-variable Krawtchouk polynomials of Rahman as well as those of Tratnik. In the two cases, the large $N$ limits of the polynomials were obtained and seen to be products of Hermite polynomials. The two sets of polynomials are eigenfunctions of the Hamiltonian and of an integral of motion whose limit is a second-order integral associated with separation of variable in Cartesian coordinates. Additionally, we have shown how these additional integrals are related by an $SU(2)$ automorphism thus providing a link between the two families of polynomials. We plan to return to this question in a future publication to provide explicit relations between Rahman and Tratnik polynomials. We have also provided models of anisotropic oscillators whose integrals can be constructed in the standard fashion from the ladder operators   
and whose wavefunctions are given by the Rahman polynomials without superfluous parameters. While we focused here, for simplicity, on two-dimensional problems, our study has a natural extension in arbitrary dimension; this shall also be the object of further investigation.

 \appendix
 
 \section{Limit of the trinomial distribution} \label{limitweight} 
Consider the limit of the trinomial distribution in the large $N$ limit, with $x$ and $y$ renormalized as in \eref{xy}. The proof follows the classical de Moivre-Laplace theorem for binomial distributions.  For notational simplicity we denote 
\be x=N\eta_1+\sqrt{N}\mu , \qquad y=N\eta_2+\sqrt{N}\nu,\label{xymu}\ee
with 
\begin{eqnarray}
 &&\mu= c_1s+c_2t, \quad \nu =c_3s +c_4t, \nonumber \\
 &&c_1c_4-c_2c_3=2\sqrt{\eta_1\eta_2(1-\eta_1-\eta_2)}.
 \end{eqnarray}
 From Sterling's formula we obtain
 \begin{eqnarray}  
W&=&{N\choose x,y} \eta_1^x\eta_2^y(1-\eta_1-\eta_2)^{N-x-y} \nonumber \\
&=&\frac{1}{2\pi}\sqrt{\frac{N}{xy(N-x-y)}}\left(\frac{x}{N\eta_1}\right)^{-x}\left(\frac{y}{N\eta_2}\right)^{-y} \nonumber \\
&&\cdot \left(\frac{N(1-\eta_1-\eta_2)}{N-x-y}\right)^{N-x-y}\left[1+\mathcal{O}(\frac{1}{N})\right].
 \end{eqnarray}
 Making the change of variables \eref{xymu}, allows for the approximation
 \be  \sqrt{\frac{N}{xy(N-x-y)}}=\frac1{N \sqrt{\eta_1\eta_2(1-\eta_1-\eta_2)}}+\mathcal{O}\left(N^{-\frac32}\right),\ee
 and hence 
 \bea W&=&\frac{1}{2\pi N \sqrt{\eta_1\eta_2(1-\eta_1-\eta_2)}}\left(1+\frac{\mu}{\sqrt{N}\eta_1}\right)^{-x} \nonumber \\
 &&\cdot \left(1+\frac{\nu}{\sqrt{N}\eta_2}\right)^{-y}\!\!\left(1-\frac{\mu +\nu}{\sqrt{N}(1-\eta_1-\eta_2)}\right)^{-(N-x-y)}+\mathcal{O}(N^{-\frac32}).\nonumber\\
  &=&\frac{1}{2\pi N \sqrt{\eta_1\eta_2(1-\eta_1-\eta_2)}}\exp\Biggl[ -x \ln\left (1+\frac{\mu}{\sqrt{N}\eta_1}\right) \nonumber \\
   && -y \ln\left (1+\frac{\nu}{\sqrt{N}\eta_2}\right)  -(N-x-y)\ln\left(1-\frac{\mu +\nu}{\sqrt{N}(1-\eta_1-\eta_2)}\right) \Biggr] \nonumber \\
 &&+\mathcal{O}(N^{-\frac32})\eea
 Finally, a Taylor series expansion of the logarithmic terms gives 
 \bea &&\left(\! \ba{c} N\\ x,y\ea\!\right)\eta_1^x\eta_2^y(1-\eta_1-\eta_2)^{N-x-y} \nonumber \\
  &&= \frac{e^{-s^2-t^2} }{2\pi N\sqrt{\eta_1\eta_2(1-\eta_2-\eta_2)}} +\mathcal{O}\left(N^{-\frac32}\right).\eea
  
 \ack {HM wishes to thank the CRM for its hospitality while this work was carried out and his work was supported by JSPS KAKENHI Grant Number 10J03343.  SP acknowledges a postdoctoral fellowship awarded by the Laboratory of Mathematical Physics of the Centre de Recherches Math\'ematiques, Universit\'e de Montr\'eal. The research of LV is supported in part by a grant from the Natural Science and Engineering Research
Council (NSERC) of Canada. A Zh wishes to thank the Centre de Recherches Math\'ematiques (CRM) at the Universit\'e de Montr\'eal for its
hospitality}

 \section*{References}
 \bibliography{all}
 \bibliographystyle{jphysa}
       
\end{document}